\documentclass[12pt,preprint]{aastex}

\usepackage{epsfig}
\usepackage{lscape}
\usepackage{natbib}
\bibpunct{(}{)}{;}{a}{}{,}
\defcitealias{ku04}{KHV04}

\newcommand{\kms}{km\,s$^{-1}$}

\shorttitle{VLA 44~GHz CH$_3$OH Masers}
\shortauthors{G\'omez et al.}
 
\begin{document}

\title{A CATALOG OF CH$_3$OH $7_0-6_1 A^+$ MASER SOURCES 
IN MASSIVE STAR-FORMING REGIONS. II.
MASERS IN  NGC~6334F, G8.67$-$0.36, AND M17} 

\author{Laura G\'omez\altaffilmark{1,2}}
\affil{Max-Planck-Institut f\"ur Radioastronomie,
              Auf dem H\"ugel 69, D-53121 Bonn, Germany}
\email{lgomez@mpifr.de}

\author{Leticia Luis, Idalia Hern\'andez-Curiel\altaffilmark{3} and 
Stan E. Kurtz}
\affil{Centro de Radioastronom\'\i{}a y 
Astrof\'\i{}sica, UNAM, Campus Morelia, Apartado Postal 3--72, 58090, Morelia,
  Michoac\'an, M\'exico}

\author{Peter Hofner}
 \affil{Physics Department, New Mexico Tech, 801 Leroy Place, 
Socorro, NM 87801, USA and National Radio Astronomy Observatory, P.O. Box 0, 
Socorro, NM 87801, USA}

\and 

\author{Esteban D. Araya}               
\affil{Physics Department, Western Illinois University, 1 University Circle, Macomb, IL 61455, USA}

\altaffiltext{1}{Centro de Radioastronom\'\i{}a y 
Astrof\'\i{}sica, UNAM, Campus Morelia, Apartado Postal 3--72, 58090, Morelia,
  Michoac\'an, M\'exico}
\altaffiltext{2}{Member of the International Max Planck
          Research School (IMPRS) for Astronomy and Astrophysics at
	  the Universities of Bonn and Cologne.}
\altaffiltext{3}{ Instituto Nacional de Astrof\'\i{}sica, \'Optica 
  y Electr\'onica, Luis Enrique Erro 1, Tonantzintla, Puebla 72840, Mexico}
  %(idaliah@inaoep.mx).

\begin{abstract}

We present Very Large Array observations of the $7_0-6_1 A^+$ methanol
maser transition at 44~GHz towards NGC~6334F, G8.67$-$0.36, and M17.
These arcsecond resolution observations complete a previous, larger
VLA survey of this maser transition in high-mass star-forming regions
reported by Kurtz et al.  We confirm the
presence of 44~GHz methanol maser emission in all three sources,
detecting eight distinct maser components in NGC~6334F, twelve
components in G8.67$-$0.36 and one in M17. 

\end{abstract}

% Keywords must be from the standard list and in alphabetical order. 

\section{Introduction}
\label{sec:intro}

Maser emission from various molecular species is a well-established
signpost of massive star formation.  Masers of the hydroxyl (OH),
water (H$_2$O), and methanol (CH$_3$OH) molecules are particularly
prevalent, and numerous studies of these masers in star formation 
regions exist in the literature; see \citet{fi07} for a recent review.

Methanol masers have sometimes proven difficult to interpret, yet they
have also been fruitful tracers of phenomena within star
formation regions \citep{el05,el06}.  
They appear in two distinct classes (I and II)
which differ in their pumping mechanisms \citep{cr92,me91a,me91b} and,
correspondingly, their locations within the star forming regions. 
Collisionally pumped class I masers are thought to be tracers of shocked gas
and hence frequently of molecular outflows 
\citep[e.g.,][]{ara10,vor10,ar09,pl90}.
Class II masers --- found in closer proximity to young
stellar objects and pumped by their mid-infrared emission --- 
show a variety of structures, including linear
\citep{mi00} and ring-like \citep{ba09} that provide 
information on the gas dynamics very close to the young stellar object
\citep{mos02}.

Methanol masers also have significant potential as tracers of magnetic field
morphology and strength via their linear polarization \citep{wie04} and
via the Zeeman effect.  Zeeman splitting has been reported for
both class I masers \citep{sa09} and class II masers
\citep{vl08,su09}.  Owing to their locations within the star-forming
regions, class I masers should be better tracers of the magnetic field
within the molecular core or clump, while class II masers should be
better tracers of the circum-protostellar magnetic field.

Quite apart from their intrinsic scientific interest, masers also
serve a valuable practical role by permitting the use of a
cross-calibration technique first described by \citet{re90};
see also the Appendix of \citet{reid97}.
The lack of nearby phase calibrators is a significant problem for
high frequency, high spatial resolution radio observations.
When sufficiently strong masers are present within the field of
view, they permit self-calibration on very short time scales. This
can substantially alleviate the problems caused by the dearth of
high frequency calibrators.
\citet{ar09}, for example, have used this technique with
44~GHz methanol masers to obtain high resolution 7~mm continuum
images of the DR21(OH) massive star formation region.

\citeauthor*{ku04} (2004; hereafter \citetalias{ku04}) presented an
arc-second resolution survey of 44 massive star formation regions in
the 44~GHz class I CH$_3$OH maser line.  At the time of their
observations, less than half of the 27 Very Large Array (VLA) antennas
were equipped with Q-band (40-50 GHz) receivers. The relatively poor
{\it uv} coverage that resulted proved problematic for imaging the
fields observed, and for three sources --- NGC~6334F, G8.67$-$0.36,
and M17 --- they were unable to uniquely determine the maser
positions, although maser emission was clearly present. More recently,
the Max-Planck Institut f\"ur Radioastronomie equipped the remaining
VLA antennas with Q-band receivers, thus mitigating the {\it uv}
coverage problems that plagued the \citetalias{ku04} survey.

The goal of the present project is to complete the \citetalias{ku04} survey by
providing accurate positions and line parameters for the 44~GHz masers
in NGC~6334F, G8.67$-$0.36, and M17.  In Sect.~\ref{sec:obs} we describe
the observations and the data reduction procedures.  In
Sect.~\ref{sec:results} we present our results, and in Sect.~\ref{sec:ind}
we provide a more detailed discussion of each source.  
A brief summary is given in Sect.~\ref{sec:conclu}.

\section{Observations and Data Reduction}
\label{sec:obs}

\subsection{Very Large Array}

NGC~6334F, G8.67$-$0.36, and M17 were observed with the Very Large
Array (VLA) of the NRAO\footnote{The National Radio Astronomy
  Observatory is operated by Associated Universities Inc. under
  cooperative agreement with the National Science Foundation.} on 2005
November 5.  We observed the methanol $7_0-6_1 A^+$ maser transition
with a rest frequency of 44\,069.43~MHz.  The array was in the D
configuration, which provides an angular resolution of about
2$\arcsec$ at 44~GHz (7~mm).  The actual resolution depends on
\textit{uv} coverage, which varies from source-to-source; the precise
resolution for each source is listed in Table~\ref{tbl-1}.  The
absolute amplitude calibrator was J1331+305 (3C286) with an adopted
flux density of 1.434 Jy at 44\,069.43~MHz. The phase calibrators were
J1700$-$261 and J1733$-$130 with bootstrapped flux densities of
0.71$\pm$0.03~Jy and 2.47$\pm$0.05~Jy, respectively.

We observed the right-hand circular polarization, using one IF and a
3.125~MHz (21 \kms) bandwidth, providing 127 spectral line channels
that were Hanning-smoothed on-line.  The 24.4~kHz  channel
widths correspond to 0.17 \kms.  Referenced pointing was performed on
the phase calibrators prior to observing the program sources. We
observed in the fast-switching mode with 160-second cycles.

The data were edited, calibrated and imaged using standard procedures
of the Astronomical Image Processing System (AIPS) of NRAO.  After the
initial external calibration, each source was imaged and inspected for
maser emission.  The brightest maser component was identified and the
peak channel was self-calibrated, first in phase, and then a second
iteration in phase and amplitude.  These self-calibration solutions
were applied to all channels, which were then imaged using weights
intermediate between natural and uniform (with the ROBUST parameter
set to 0) and CLEANed in an iterative fashion.  For the initial
iteration, clean boxes were assigned only to the strongest masers; for
subsequent iterations additional clean boxes were added, as weaker
masers became visible.  A final image cube was made for each source,
with clean boxes for all identified maser components, and CLEANed to a
level of twice the theoretical \textit{rms}.

The maser parameters (see Table 2) were extracted from the final image
cubes using the AIPS tasks JMFIT, IMSTAT and ISPEC.  The absolute flux
calibration uncertainty is $\sim15\%$ and 
we estimate the absolute positional uncertainty to be $0\farcs2$ for all
the masers, although the uncertainty of the Gaussian fit of stronger
masers is smaller \citep{re88}."

\subsection{Spitzer Space Telescope}
\textit{Spitzer} images, shown in figures 1--3, were taken from the 
Galactic Legacy Infrared Mid-Plane Survey Extraordinaire 
\citep[GLIMPSE;][]{be03} program, based on observations with the IRAC camera 
\citep{fa04}.

\section{Results} \label{sec:results} 
We confirm the presence of 44~GHz CH$_3$OH
maser emission in all three sources, detecting eight distinct maser
components in NGC~6334F, twelve masers in G8.67$-$0.36 and one maser in
M17. 

The observed parameters of all detected masers are listed in
Table~\ref{tbl-2}. Column (1) gives the source name, while columns (2)
and (3) give the J2000 peak position, determined from a
two-dimensional Gaussian fit to the peak channel.  Column (4) gives
the peak flux density, also from the Gaussian fit.  Column (5) gives
the LSR-velocity of the peak channel, while column (6) provides the
the FWZI (full width at zero intensity) at the 3$\sigma$ level.  If
multiple velocity components at the same sky position are present then
we report the full velocity range, even if some intermediate channels
fall below the 3$\sigma$ level. Column (7) gives the integrated line
flux, calculated as $\Sigma$S$_i$\,$\Delta$V$_i$ summed over channels
above the 3$\sigma$ level.

\begin{figure}
\epsscale{.80}
  \plotone{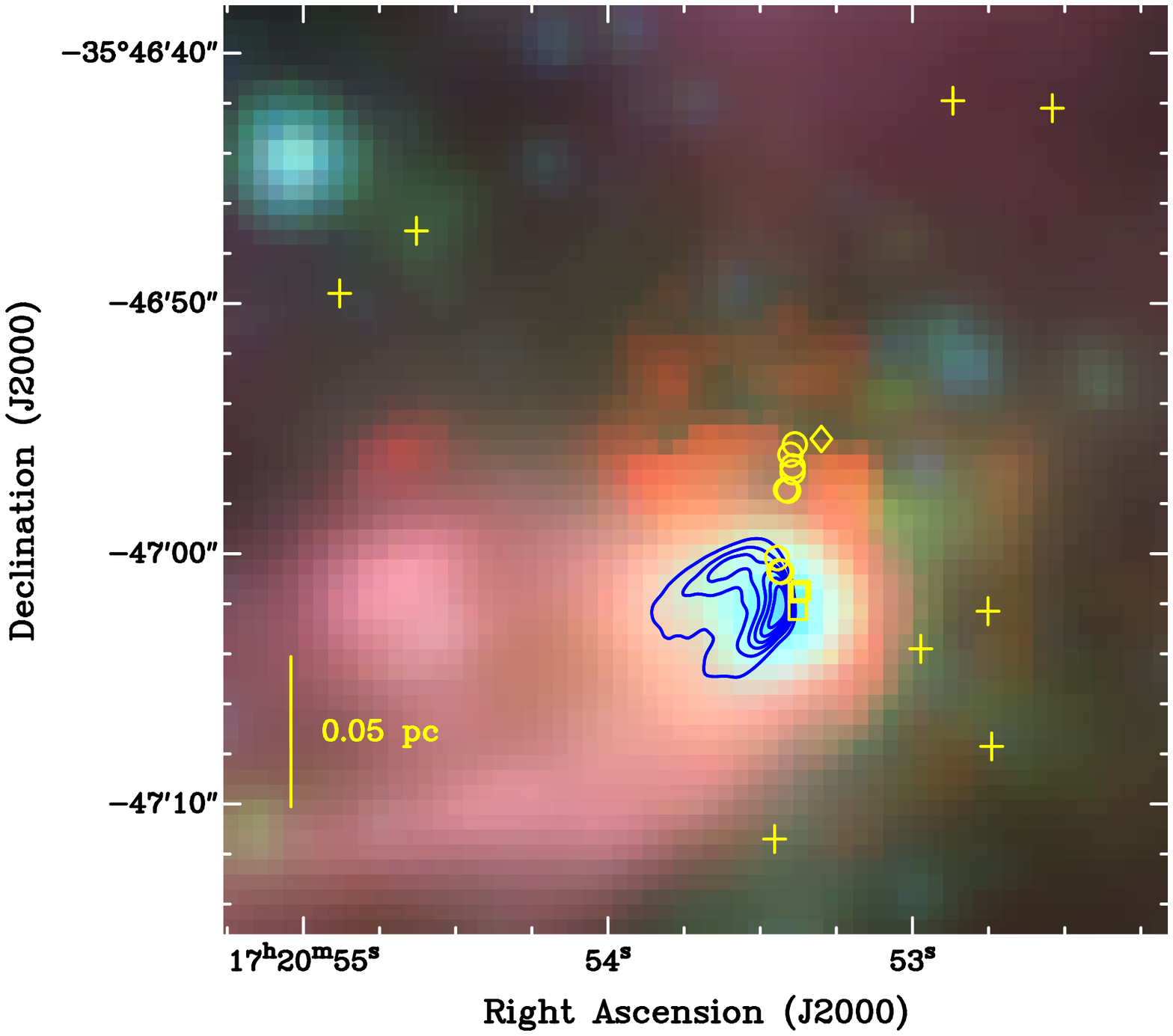}
  \caption{
VLA 3.6 cm continuum contours \citep{ca02}
overlaid on the three-color GLIMPSE IRAC image of NGC~6334F
showing 8 $\mu$m (red), 4.5 $\mu$m (green), and 3.6 $\mu$m (blue)
emission. The contour levels are from 10\% to 85\% (step 15\%) of the 
peak emission of 127.5 mJy beam$^{-1}$. ``Plus'' symbols represent 44~GHz 
CH$_3$OH masers (this work), while squares indicate OH masers \citep{br01},
circles represent H$_2$O masers \citep{fo89}
 and the diamond indicates a 23~GHz methanol maser \citepalias{ku04}.
 \label{fig1}}
\end{figure}

\begin{figure}
\epsscale{.80}
\plotone{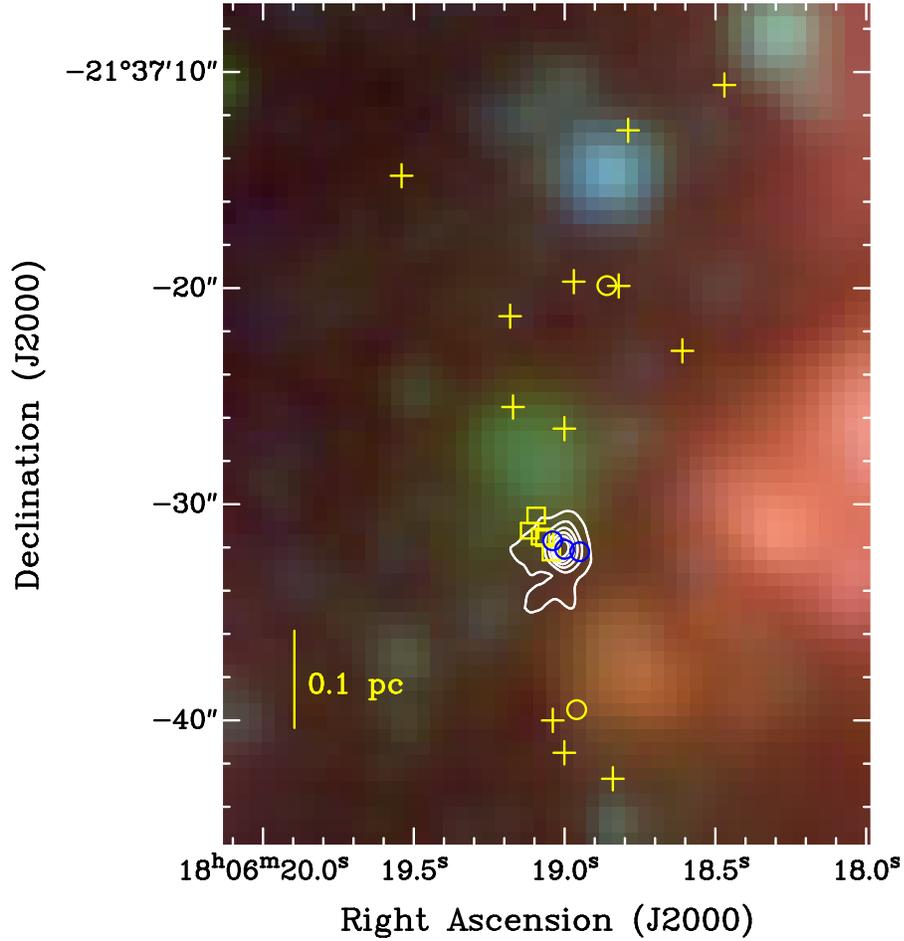}
  \caption{
VLA 2 cm continuum contours \citep{wo89}
overlaid on the three-color \textit{Spitzer}/GLIMPSE image of G8.67$-$0.36
showing 8 $\mu$m (red), 4.5 $\mu$m (green), and 3.6 $\mu$m (blue) emission.
The contour levels are from 10\% to 85\% (step 15\%) of the 
peak emission of 124.8 mJy beam$^{-1}$.
``Plus'' symbols represent 44 GHz CH$_3$OH masers (this work),
while squares indicate OH masers \citep{fo89} and
 circles indicate H$_2$O masers  \citep{ho96}.\label{fig2}}
\end{figure}

\begin{figure}
\epsscale{.80}
\plotone{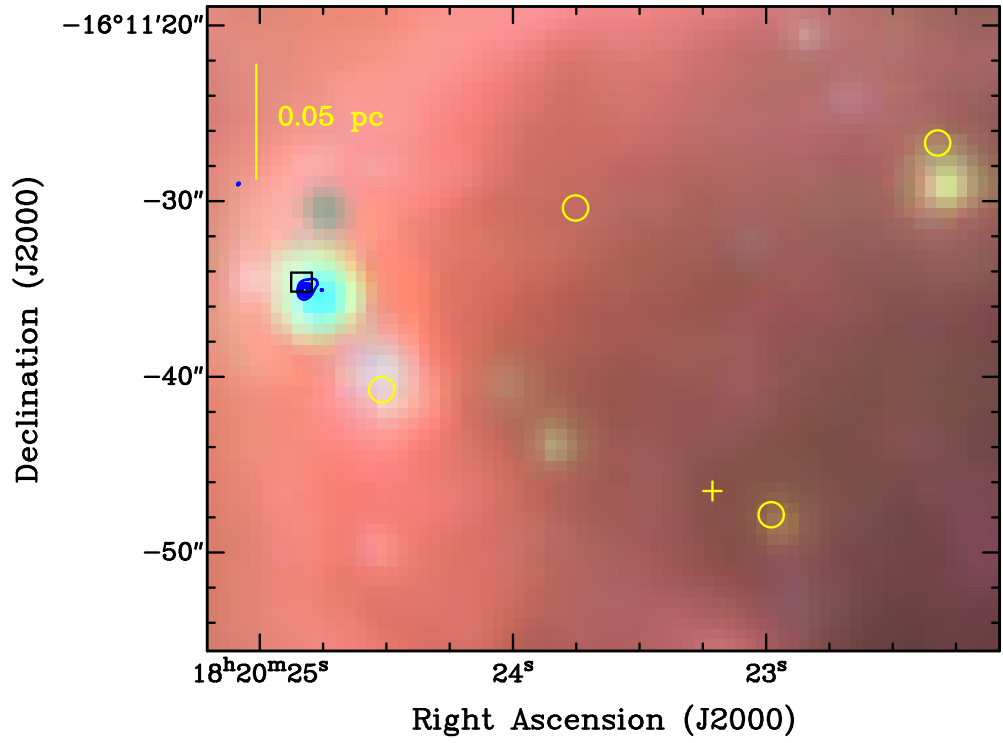}
  \caption{
VLA 2 cm continuum contours \citep{wo89}
overlaid on the three-color \textit{Spitzer}/GLIMPSE image of M17
showing 8 $\mu$m (red), 4.5 $\mu$m (green), and 3.6 $\mu$m (blue) emission. 
The contour levels are from 5\% to 85\% (step 20\%) of the 
peak emission of 105.6 mJy beam$^{-1}$.
The ``plus'' symbol shows the position of the 
 44~GHz CH$_3$OH maser (this work), the square represents an OH maser 
\citep{brog01}, and circles indicate H$_2$O masers \citep{jo98}.\label{fig3}}
\end{figure}

Figures~\ref{fig1}--\ref{fig3} show the three-color GLIMPSE image of
each region, with radio continuum emission plotted as contours and
various maser species shown as symbols.  The symbol sizes are larger
than the positional uncertainty in all cases.

\section{Discussion of Individual Sources}
\label{sec:ind}

\subsection{NGC~6334F}

\label{sec:ngc}

NGC~6334F is a well-known UC HII region, lying within the NGC~6334
cloud complex, at a distance of 1.7 kpc \citep{ne78}. It is also known
as NGC~6334I, where ``I'' is a roman numeral ``one'', originating from
far-infrared studies \citep[e.g.,][]{ge82}.  We adopt the convention
of \citet{ro82}, in which the letter ``F'' refers to the UC HII
region.

The eight 44~GHz class I masers detected within the one arcminute VLA
primary beam indicate a higher level of maser activity than the
majority of the sources in the \citetalias{ku04} survey, which has a
median of four maser features per field.  Figure~\ref{fig1} shows the
three-color \textit{Spitzer}/GLIMPSE image of NGC~6434F and the
location of the maser components.  The eight masers are distributed
over an area $\sim$ 0.25 pc $\times$ 0.25 pc. As is typical for class
I masers, these eight components do not appear to be associated with
the UC HII region, other masers, or the IR emission.  Nor do they 
coincide with thermal ammonia peaks or ammonia masers as reported
by \citet{beu05,beu07} or with the millimeter peaks reported by
\citet{hu06}.  The average projected distance from the masers to the
geometric center of the UC HII region is 0.12 pc.

Neglecting the two northern-most masers, there is a southwest
to northeast positional orientation of the remaining six masers.  This
orientation corresponds to the blue-shifted (southwest) and
red-shifted (northeast) high velocity outflow mapped with the APEX 12
m telescope by \citet{le06}. There is a weak tendency in the maser
velocity structure in accordance with this pattern: the average
velocity of the southwestern masers is $-9.5$~\kms~while the
average for the northeastern masers is $-6.3$~\kms.  We suggest
that these six masers  are related to the bipolar outflow
reported by \citet{le06}.  We caution, however, that the trends in
both position and velocity are not particularly strong.

The velocity range of emission that we detected ($-10.9$ to $-4.6$
\kms) is slightly shifted from that reported by \citetalias{ku04}
($-9.0$ to $-2.5$ \kms) and also differs from the single-dish
observations of \citet{sl94} ($-8.4$ to $-4.8$ \kms).

\subsection{G8.67$-$0.36}
\label{sec:g8}

The UC HII region G8.67$-$0.36,  at a distance of
4.8~kpc \citep{fi03}, was classified by \citet{wo89} as having a
core-halo morphology; i.e., a single compact peak surrounded by an
extended, low-surface-brightness halo.

As in NGC~6334F, the relatively large number of masers detected in
this field (12) indicates an unusually high level of maser activity.
We identify two regions of maser activity in the field (see
Fig.~\ref{fig2}): one to the north and the other to the south of the
UC HII region.  No velocity trend with respect to position is seen.
The strongest two masers in this field lie at the edge of an Extended Green
Object \citep[EGO;][]{cy08}, consistent with the idea that EGOs trace
molecular outflows from MYSOs and that class I methanol masers arise
from the interaction of outflows with dense clumps of gas
\citep{pl90}.  \citet{cy08} did not catalog the G8.67$-$0.36 region
because it lies outside their survey area ($10\arcdeg < l <
65\arcdeg$ and $295\arcdeg < l < 350\arcdeg, b = \pm 1\arcdeg$).
Although eight of the twelve masers lie at the edge of some infrared
feature, four of the masers appear relatively isolated from the
infrared emission; hence our data do not suggest a unique correlation
of these masers with a particular set of gas conditions.

\citetalias{ku04} report maser emission at from 33 to 37~\kms, which
is in close agreement with the velocity of HCO$^+$ and H$^{13}$CO$^+$
emission \citep[34.8 \kms;][]{pu06}. We detect maser emission in a
velocity range of 33.6--39.9 \kms.   

The average projected distance from the masers to the geometric 
center of the UC HII region is 0.29 pc.

\subsection{M17}
\label{sec:m17}

The M17 region hosts, among other features, the cometary UC
HII region UC-1 \citep[e.g.,][]{fe80, jo98}.  Distances
reported for the M17 nebula have ranged from 2.2 kpc \citep{ch80} to
1.3 kpc \citep{ha97}. More recently, a distance of 1.6 kpc has been
reported by \citet{ni01}, which we adopt here.

Unlike the previous two sources, M17 presents very limited 44 GHz
maser activity with only a single maser component detected in the
field, at a projected distance of 0.2 pc from the UC HII region (see
Fig.~\ref{fig3}).  The maser properties are listed in
Table~\ref{tbl-2}; the 19.1 \kms~velocity that we find is
the same as found by \citetalias{ku04}.

\section{Summary}
\label{sec:conclu}

Using the Very Large Array, we have observed 44~GHz class I methanol
maser emission in the massive star-forming regions NGC~6334F,
G8.67$-$0.36, and M17.  Our principal result is to provide accurate
maser positions and parameters, thus completing the catalog of
\citetalias{ku04}.

In addition, we find that: (1) Two of the sources (NGC~6334F and
G8.67$-$0.36) show significantly higher levels of maser activity than
the typical survey source. 
(2) For all three sources the masers are well-separated from the HII
region, with projected distances ranging from 0.1 to 0.3 pc.  This is
in good agreement with the \citetalias{ku04} survey, which found a median
separation of 0.2 pc for a subsample of 22 sources that had both HII
regions and maser emission.

\acknowledgments

We thank Prof. K. M. Menten for a very helpful review of the manuscript. 
We are grateful to the National Radio Astronomy Observatory for making
these observations possible through their program of observing time for
university classes. This work is based in part on observations made with
the \textit{Spitzer Space Telescope}, which is operated by the Jet Propulsion
Laboratory, California Institute of Technology under a contract with NASA.
This research has made use of of NASA`s Astrophysics
Data System Bibliographic Services and the SIMBAD database operated at CDS, 
Strasbourg, France.
L. G.,  L. L. and I. H. acknowledge the support of CONACyT,
M\'exico. L. G. was supported for this research through a stipend from the
 International Max Planck Research School (IMPRS) for Astronomy and
Astrophysics at the Universities of Bonn and Cologne.
S. K. acknowledges partial support from UNAM-DGAPA grant IN101310.
P. H. acknowledges partial support from NSF grant AST-0908901.
\bibliographystyle{apj} % style rmaa.bst
\bibliography{masersb}

\clearpage

\clearpage

\clearpage

\clearpage

\begin{table}
\small
\begin{center}
\caption{Observed Sources.\label{tbl-1}}
\begin{tabular}{lccccc}
\tableline\tableline
Source &\multicolumn{2}{c}{Pointing Center\tablenotemark{a}} & Central Velocity & Synthesized Beam\tablenotemark{b}& Channel Map rms\\
\cline{2-3} 
 &  R.A.(J2000) & Dec(J2000) & (\kms) &   & (mJy beam$^{-1}$)\\ 
\tableline
NGC~6334F & 17 20 54.00 & $-35$ 47 00.0&$-10.4$& $3\farcs 83 \times 1\farcs 27$; $-2\arcdeg$ &45\\ 
G8.67$-$0.36& 18 06 19.20&$-21$ 37 30.0&$+30.6$& $2\farcs 41\times1\farcs 31$; $-7\arcdeg$&40\\
M17 & 18 20 24.40 & $-16$ 11 32.0 & $+14.0$&$2\farcs 07 \times 1\farcs 36$; $-10\arcdeg$& 47   \\ 
\tableline
\end{tabular}
\tablenotetext{a}{Units of right ascension are hours, minutes, and seconds, 
 and units of declination are degrees, arcminutes, and arcseconds.
 }
\tablenotetext{b}{Major axis $\times$ minor axis; position angle of major axis.}
\end{center}
\end{table}

\clearpage

\begin{table}
\begin{center}
\caption{44~GHz Maser Parameters.\label{tbl-2}}
\begin{tabular}{lcccccc}
\tableline\tableline
Source &\multicolumn{2}{c}{Maser Peak Position\tablenotemark{a}} & $S_{\rm Peak}
$ 
 & V$_{\mathrm {LSR}}$ & $\Delta$V\tablenotemark{b} & $\int S\, d\mathrm{V}$ \\
\cline{2-3} 
&  R.A. (J2000) & Decl. (J2000) & (Jy) & (\kms) & (\kms) & (Jy \kms) \\ 
(1)&(2)&(3)&(4)&(5)&(6)&(7)\\
\tableline
    NGC~6334F 
& 17 20 52.54& $-35$ 46 42.2&  1.3& $-8.1$ & 1.5                 & 1.1 \\
& 17 20 52.74& $-35$ 47 07.7&  1.2& $-10.2$& $-10.9$ to $-8.7$   & 1.0 \\
& 17 20 52.75& $-35$ 47 02.3&  5.9& $-6.2$ & 1.3                 & 4.1 \\
& 17 20 52.87& $-35$ 46 41.9&  1.5& $-8.1$ & $-9.0$ to $-6.9$    & 1.1  \\
& 17 20 52.97& $-35$ 47 03.8&  2.9& $-8.9$ & $-10.9$ to $-6.2$   & 2.7  \\
& 17 20 53.45& $-35$ 47 11.4&  0.7& $-10.1$& 1.7                 & 0.6 \\
& 17 20 54.63& $-35$ 46 47.1&  9.4& $-6.9$ & $-7.4$ to $-4.6$    &12.6 \\
& 17 20 54.88& $-35$ 46 49.6&  2.8& $-5.7$ & 1.2                 & 1.1 \\
G8.67$-$0.36                                                    
& 18 06 18.47& $-21$ 37 10.6&  0.4&  37.9  & 0.5                 & 0.1 \\
& 18 06 18.61& $-21$ 37 22.9&  1.9&  38.1  & 0.7                 & 0.7 \\
& 18 06 18.79& $-21$ 37 12.7&  1.3&  37.7  & 1.3                 & 0.9 \\
& 18 06 18.82& $-21$ 37 19.9&  1.3&  38.7  & 36.9 to 39.9        & 0.8 \\
& 18 06 18.84& $-21$ 37 42.7&  1.4&  34.8  & 1.2                 & 0.7\\
& 18 06 18.97& $-21$ 37 19.7&  6.2&  37.6  & 36.9 to 38.7        & 4.2 \\ %
& 18 06 19.00& $-21$ 37 41.5&  7.9&  35.1  & 1.3\tablenotemark{c}& 5.3\tablenotemark{c}\\ %
& 18 06 19.00& $-21$ 37 26.5& 16.5&  35.4  & 34.1 to 38.2        &21.1 \\ %
& 18 06 19.04& $-21$ 37 40.0&  9.3&  33.6  & 1.8\tablenotemark{c}& 9.6\tablenotemark{c}\\ %
& 18 06 19.17& $-21$ 37 25.5& 18.2&  35.9  & 2.5                 &13.2 \\
& 18 06 19.18& $-21$ 37 21.3&  1.4&  33.8  & 2.2                 & 1.1 \\
& 18 06 19.54& $-21$ 37 14.8&  0.9&  36.4  & 35.0 to 36.9        & 0.7 \\
M17  
& 18 20 23.21& $-$16 11 46.5& 9.0 &  19.1  &  0.7                & 0.5 \\ 
\tableline
\end{tabular}
\tablenotetext{a}{Units of right ascension are hours, minutes, and seconds, and 
units of declination are degrees, arcminutes, and arcseconds.}
\tablenotetext{b}{A single number indicates the line width for $S > 3\sigma$.
Two numbers indicates the velocity range when a single
sky position has multiple components, even if some intermediate channels
fall below 3$\sigma$.}
\tablenotetext{c}{These two masers are spatially blended;
$\Delta$V and $\int S\,d\mathrm{V}$ are approximate values.}
\end{center}
\end{table}

\end{document}